\documentclass{article}

\PassOptionsToPackage{numbers, compress}{natbib}

\usepackage[final]{nips_2018} 

\usepackage[utf8]{inputenc} 
\usepackage[T1]{fontenc}    
\usepackage{hyperref}       
\usepackage{url}            
\usepackage{booktabs}       
\usepackage{amsfonts}       
\usepackage{nicefrac}       
\usepackage{microtype}      

\usepackage{graphicx}
\usepackage{amsmath}
\usepackage{amssymb}

\title{Minority report detection in refugee-authored community-driven journalism using RBMs}

\author{
  Bogdana Rakova \\
  Think Tank Team\\ 
  Samsung Research America\\
  665 Clyde Ave, Mountain View, CA\\
  \texttt{b.rakova@samsung.com} \\
  \And
  Nick DePalma \\
  Artificial Intelligence Center\\ 
  Samsung Research America\\
  665 Clyde Ave, Mountain View, CA\\
  \texttt{n.depalma@samsung.com}\\
}

\begin{document}

\maketitle

\begin{abstract}
Our work seeks to gather and distribute sensitive information from refugee settlements to stakeholders to help shape policy and help guide action networks. 
In this paper, we propose the following 1) a method of data collection through stakeholder organizations experienced in working with displaced and refugee communities, 2) a method of topic modeling based on Deep Boltzmann Machines that identifies topics and issues of interest within the population, to help enable mapping of human rights violations, and 3) a secondary analysis component that will use the probability of fit to isolate minority reports within these stories using anomaly detection techniques. 
\end{abstract}

\section{Introduction}
The United Nations Sustainable Development Goals, or SDGs\cite{sdgs2015united}, are a set of seventeen objectives that must be addressed in order to bring the world to a more equitable, prosperous, and sustainable path. 
Our work is focused on how new types of AI tools could help promote peaceful and inclusive societies for sustainable development, provide access to justice for all and build effective, accountable and inclusive institutions at all levels\cite{sdgs2015united16}. 
In our analysis and proposal we focus on those fleeing violence, turmoil, or seeking opportunity in a foreign country where they are often denied their basic human rights and liberties\cite{h12}. 

Expressive journaling techniques have frequently been employed to moderate strong emotional experience and to aid in coping\cite{niles2014randomized}. Large N trials have demonstrated that journaling can help high anxiety patients lower blood pressure, reduce recurring anxious thoughts, and cope with traumatic experience. While we do not expect these effects to compensate for the experiences refugees go through, we do believe that allowing those refugees to express themselves in a safe environment is mutually beneficial to all stakeholders.
Our hypothesis is that, \textit{many journal entries will share subject matter, major events, and potential solutions} within the settlement. Using topic modelling, we can track these frequently told stories. But while major events like publicly viewed human rights violations may be frequently clustered to a topic, lesser and more intimate violations such as sexual abuse may not be modeled as well in standard topic models. 


We present an unsupervised study of journal articles using Deep Boltzmann Machines(DBMs)\cite{pmlr-v5-salakhutdinov09a}. 
We analyze the potential implications of the development of community-driven journalism tools in the broader ecosystems of organizations and individuals working to provide sustainable solutions to the problems facing displaced communities. We discuss foreseeable challenges, threat models and next steps. 


\section{Related work}
Making long lasting impact requires coordination with organizations with impact and reach to the populations we are most interested in. Microsoft's Project Fortis\cite{schlegel2017} is one such project that  dynamically aggregates data and enables data scientists to perform statistical analysis collected from popular social media and humanitarian databases like those published from OCHA\cite{nations2018}. We differ from this system in that we are interested in proposing models that identify minority reports using meta analysis of documents originating from refugee settlements. 
Rather than build tools to aggregate data, we hope to use AI techniques to reveal patterns and highlight minority reports that need attention and resources.



Deep Learning has recently been applied in the context of coordinating humanitarian relief efforts in conflict situations through the analysis of remote sensing imagery from satellites or drones\cite{Quinn20170363}. Our work develops a complimentary approach that highlights the importance of community authored stories. Topic modeling is a key component in the fields of information retrieval\cite{Wei:2006:LDM:1148170.1148204} and understanding\cite{Hall:2008:SHI:1613715.1613763}.

We build on the analysis of the main challenges posed by Hu et al.\cite{Hu2014} and focus on the way non-experts can use topic modeling tools to aid their needs.  We differ from their system as our topic modeling approach uses a different methodology that potentially would allow the system to reason about anomalies as well as analyze multimodal data\cite{Andrzejewski:2009:IDK:1553374.1553378}.
 

\section{Story collection}
The goal of our work is to analyze if AI-enabled topic modeling and anomaly detection can aid the work of humanitarian action and advocacy groups, and investigative journalists\cite{propublica} by helping them organize different pieces of evidence. We begin by acknowledging that a transparent data collection process is crucial for the real-world success of any potential proposal. It must preserve peoples' privacy and must be informed by all involved stakeholders. For the purpose of our experiments, we use a database consisting of 6,258 news articles published by major media outlets between 01/2016 - 09/2017 and mentioning the refugee crisis\cite{kaggleAllTheNews}. In future work we aim to collaborate directly with practitioners\cite{velosyouth, advocatedAbroad, campfire, refugeeJournalism} to help them extract and map insights from case work related to human rights violations in refugee settlements.   


\section{Methodology and preliminary results}
\begin{figure}
  \centering
  \fbox{
  \rule[-.5cm]{0cm}{4cm} 
  \includegraphics[width=0.4\textwidth]{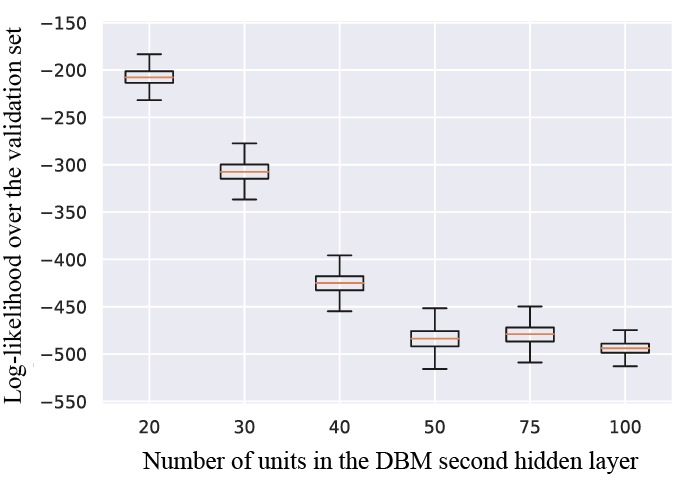}
  \hspace*{1cm}
  \includegraphics[width=0.4\textwidth]{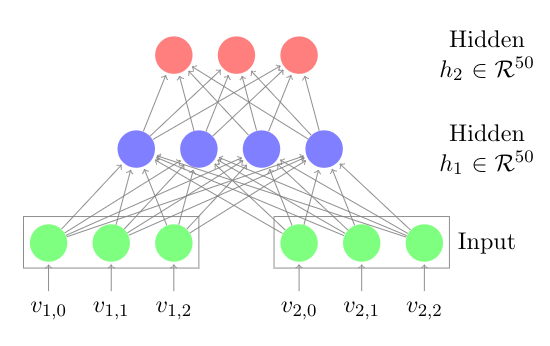} 
  \rule[-.5cm]{0cm}{4cm}
  }
  \caption{DBM hidden layer architecture analysis(left) and our Replicated Softmax RBM per document model(right). }
  \label{fig:DBM}
\end{figure}

We start by training a Deep Boltzmann Machine model (DBM) - a two-hidden layer Replicated Softmax RBM\cite{NIPS2009_3856} with weights sharing. The visible units in our architecture correspond to lemmatized word count vectors where each vector represents a document in the training corpus(see Figure \ref{fig:DBM}, right). As with a standard RBMs the learning proceeds via Contrastive Divergence\cite{hinton2002training}.

We hypothesize that the DBM model learns the peculiarities that are more prevalent in the training data, producing better reconstructions of them. We hypothesize that minority report entries will occur rarely in the provided training data, preventing the DBM from learning and reconstructing it. Therefore we develop a metric for discovering these minority reports by measuring the distance between a document's input vector and its reconstruction. We aim to highlight those documents so that social scientists interacting with the algorithmic system could leverage these information retrieval insights. 
Following training of the traditional DBM parameters, $W_{1,i}, W_2,b_{1,i},b_2$ corresponding to the weights and biases found for document $i$, we compute the reconstruction error. We first define the forward-backward pass mapping as follows: 

\begin{align}
\tilde{v}_i = \sigma(W_ 2 \cdot \sigma( W_1 \cdot  v_i  + b_ 1) + b_ 2) \\
\hat{v}_i = \sigma(W_1 \cdot \sigma(\tilde{v}_i \cdot W_2 + b_ 2') + b_1') \\
\epsilon(v_i, \hat{v}_i) = |\hat{v}_i-v_i|
\end{align}

Where $v_i$ is the word count vector from a predefined dictionary for a specific document $i$, $\tilde{v}_i$ is the latent hidden embedding of that document from the DBM and $\hat{v}_i$ is the reconstruction of $v_i$ given $\tilde{v}_i$. Finally $\sigma$ is the non-linear activation function, $\sigma(x)={1+\tanh({x \over 2}) \over 2}$. The dictionary consisted of the top 1000 most common terms in the training corpus. The weights and biases represent the symmetric interaction terms between visible-to-hidden and hidden-to-hidden variables learned while pre-training the Softmax RBMs and fine-tuning the DBM by minimizing the negative log-likelihood over a holdout validation set(Figure \ref{fig:DBM}, left). 

\begin{figure}
  \centering
  \fbox{
  \includegraphics[width=0.9\textwidth]{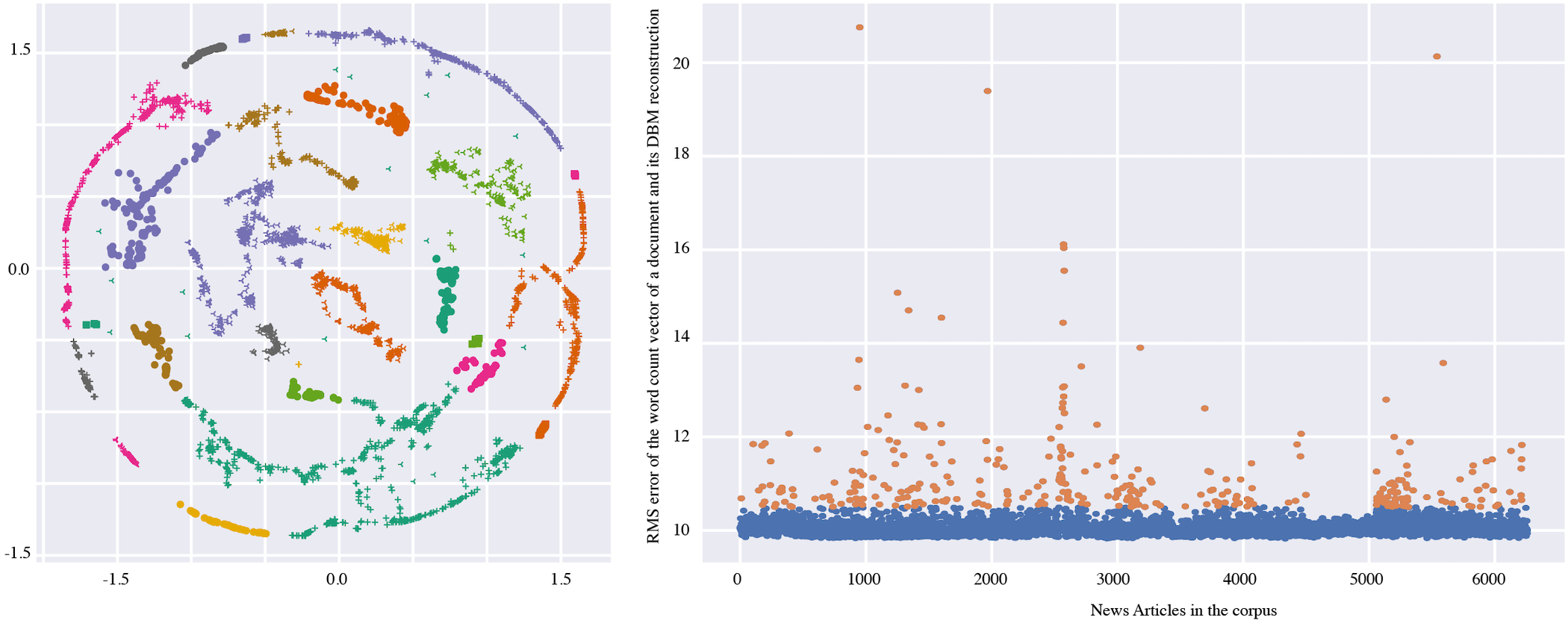}
  }
  \caption{The results from applying DBSCAN clustering of the DBM representations of all articles in the corpus(left) and minority report selection using our metric. Selected minority reports are highlighted in orange(right).}
\label{fig:tsne}
\end{figure}

We have performed an initial investigation into our trained model by first generating the latent embeddings $\tilde{v}$ of all the documents in the corpus and clustering the results using DBSCAN, an agglomerative clustering technique. 


We visualized the results in Figure \ref{fig:tsne}, left using a t-SNE visualization\cite{maaten2008visualizing}. By analyzing word frequencies in each cluster we find that for more than half of all clusters the US President Trump is a major actor, however individual subclusters were formed related to other political figures such as the US Secretary of State John Kerry and Germany's Chancellor Angela Merkel. Another subcluster emerges related to terms such as women, public health and policy.
Finally, we use our minority report metric to extract the distance between a document and its reconstruction and plot the result in Figure \ref{fig:tsne}, right. As discussed previously, we believe that the outliers indicate that some of these documents were not modeled well and that their reconstruction error would be significantly larger than the other documents in the corpus. 

Initial probing of these minority reports show that they cover the meta-topics of: 1) articles not related to the refugee crisis at all, 2) articles expressing specific sentiment which we don’t see often in media, 3) articles which were very concrete and graphic about violence, 4) articles about the Cold War, and 5) an assortment of other topics.


\section{Challenges and future work}
Violent conflicts and natural disasters are causing large numbers of civilian casualties\cite{sdgs2015united16}. Communities bring the energy and expertise to reinvent themselves from within. They know what they need more than anybody that we could possibly bring in from the outside. Algorithmic tools could aid multi-lingual cross-cultural understanding however we need a broader conversation where vulnerable communities are empowered to participate.

We understand that sensitive data in refugee settlements can be used for ill purposes and would like to highlight the importance of future work in the fields of privacy, trust and secure information sharing systems. Loss, theft, abuse, misuse, and unintended actions with datasets threaten the lives of these individuals and may lead to irreversible consequences\cite{harvard2018}. 

It is important to consider that any algorithmic tool we create will be used in unintended ways and therefore we need to create threat models that identify vulnerabilities and define countermeasures to prevent, or mitigate the effects of threats. Through conversations with stakeholders, we have also found that the expected number of documents are far fewer than the dataset that was used for our experiment. Future work should explore how to perform the same type of analysis with fewer examples. Furthermore, could technology help us detect artificially crafted and fake data designed to make people believe and act in certain ways? We need to ask these questions and rigorously study the implications of the available technological tools in the broader social context where they are being applied.

\section{Acknowledgements}

This work was greatly influenced by Brent Dixon, co-organizer and chair of Greece Communitere - the Greece chapter of an international NGO creating dynamic, collaborative hubs in displaced and post-disaster communities. We would also like to thank the generosity of Samsung Research for encouraging our investigation into these issues.

\bibliographystyle{ieeetr}
\bibliography{library}

\end{document}